\documentstyle[osa,twocolumn]{revtex}

\begin{document}

\preprint{submitted to Optics Letters}

\title{Spontaneous emission spectrum of the non-lasing supermodes in 
       semiconductor laser arrays}

\author{Holger F. Hofmann and Ortwin Hess}
\address{Institute of Technical Physics, DLR\\
Pfaffenwaldring 38--40, D--70569 Stuttgart, Germany}

\date{\today}

\maketitle
\begin{abstract}
It is shown that the interference between spontaneous emission into the non-lasing supermode 
and the laser field of a semiconductor laser array causes spatial holeburning which couples
the dynamics of the spontaneous emission with the laser field. 
In particular, phase locking between the spontaneous emission and the lasing mode leads to
the formation of a spectral triplet composed of in-phase relaxation oscillation
sidebands and an out-of-phase line at the lasing frequency. 
\end{abstract}

\vspace{0.5cm}


The investigation of spontaneous emission into the non-lasing modes of
a semiconductor laser operating in cw-mode can provide valuable insights into the 
carrier dynamics of the laser device \cite{Hof97,Jan97}. 
This is especially useful in the study of phase-locked laser arrays, 
such as two-dimensional arrays of vertical
cavity surface emitting lasers (VCSEL arrays) \cite{Cat96,Mor93,Mor92,Ore92}. 
Such coherently coupled VCSEL arrays are fabricated e.g.~to obtain maximal coherent output power. 
They usually exhibit stable anti-phase locking between
adjacent lasers resulting in the lasing of only the supermode with highest frequency \cite{Cat96,Mor93,Mor92,Ore92}. 

However, there is still spontaneous emission into the other non-lasing modes \cite{Mor92}.
Generally, the interference terms between the laser light
and the non-lasing modes determine the spatial intensity distribution inside the laser cavity. 
The significance of these terms depends on the phase relation between the spontaneous
emission and the laser light. It is particularly strong if the spontaneous
emission is in phase with the laser light.  
Consequently one can expect that the in-phase spontaneous emission
will cause spatial hole-burning and spatial relaxation oscillations,
while the out-of-phase spontaneous emission will not couple to  
the carrier dynamics. 
 
In this letter, we describe the dynamics of the lasing antisymmetric
supermode and the non-lasing symmetric supermode of a two laser array
using a rate equation
model similar to the one introduced by Winful and Wang \cite{Wan88,Win88}.
In this model, the spatial distribution of the carrier density is 
approximated by assigning separate carrier densities to the individual 
lasers of the array. For a two laser array, the dynamical equations of the
field and the carrier densities may be written in terms of the 
anti-symmetric supermode $E_-$ and the symmetric supermode $E_+$.
The validity of this discretization of the electromagnetic field and the limitation to two modes
is established in a more detailed model based on partial differential equations \cite{Mun97}. 
The total carrier density $N$, which is the sum of both carrier densities in the lasers,
interacts equally strong with both modes. The spatial carrier density 
difference between the two lasers, $\Delta$, is a measure of spatial 
hole-burning effects on the length scale of the array. The dynamical equations 
then read
\begin{mathletters}
\begin{eqnarray}  
\frac{d}{dt} E_+ &=& 
 \frac{w}{2}   N  (1-i\alpha)E_+ -(\kappa_++i\omega_+)E_+ \nonumber \\ &&
+\frac{w}{2}\Delta(1-i\alpha)E_- \\  
\frac{d}{dt} E_- &=& 
 \frac{w}{2}   N  (1-i\alpha)E_- -(\kappa_-+i\omega_-)E_- \nonumber \\ &&
+\frac{w}{2}\Delta(1-i\alpha)E_+ \\  
\frac{d}{dt} N &=& \mu - \gamma N - w(E_+^*E_++E_-^*E_-)   N  \nonumber \\ &&
                                  - w(E_+^*E_-+E_-^*E_+)\Delta \\
\frac{d}{dt} \Delta &=& -(\gamma+2\Gamma)\Delta - w(E_+^*E_++E_-^*E_-)\Delta
                                \nonumber \\ && - w(E_+^*E_-+E_-^*E_+)   N.
\end{eqnarray}
\end{mathletters}
The linear optical properties
of the laser cavity are given by the loss rates $\kappa_+$ and $\kappa_-$ 
and the frequencies at $N=0$, $\omega_+$ and $\omega_-$. The symmetric 
supermode $E_+$ and the anti-symmetric supermode $E_-$ are eigenmodes of the 
linear optical equations.
The optical coupling between the two lasers is given by the differences 
between the eigenvalues, $s = \kappa_+ - \kappa_-$ and $\Omega = \omega_-
-\omega_+$. 
The gain and the carrier induced frequency shift are represented by the
linear gain coefficient $w/2$ and the linewidth enhancement factor $\alpha$, respectively.
$w$ is the linear gain coefficient of the individual laser. The factor of
$1/2$ is a result of the normalization of the electromagnetic modes in the
two laser array.

The timescales of the carrier dynamics are given by the injection rate
above transparency
$\mu$, the recombination rate at zero field $\gamma$ and a diffusion
rate $\Gamma$. The diffusion rate $\Gamma$ may be derived from the 
ambipolar diffusion constant of the carriers $D_{\mbox{diff}}$ by solving
the diffusion equation for a carrier density distribution modulated with a
period equal to the distance $r$ between the lasers in the array. The 
diffusion rate is thus 
\begin{equation}
\Gamma = \frac{4\pi^2}{r^2}D_{\mbox{diff}}.
\end{equation}  
Note that such a treatment of diffusion represents an extension of the original
model of Winful and Wang \cite{Wan88,Win88}, as discussed elsewhere~\cite{Hof98}.

The stable solution of the array dynamics is given by $\Delta=0$, 
$N=2\kappa_-/w$, $E_+=0$ and 
$E_-=\sqrt{I_0} \exp [-i(\omega_-+\alpha\kappa_-)t]$, where the laser 
intensity $I_0$ is a linear function of the injection rate $\mu$.
The linearized dynamics of small fluctuations in the non-lasing mode $E_+$
and the spatial hole-burning parameter $\Delta$ read
\begin{mathletters}
\begin{eqnarray}  
\frac{d}{dt} E_+ &=& 
 -(\kappa_+-\kappa_-+i\omega_++i\alpha\kappa_-)E_+ \nonumber \\ &&
+\frac{w}{2}\Delta(1-i\alpha)\sqrt{I_0}\; e^{-i(\omega_-+\alpha\kappa_-)t}\\  
\frac{d}{dt} \Delta &=& -(\gamma+2\Gamma+wI_0)\Delta \nonumber \\&&
          - 2\kappa_-\sqrt{I_0}(e^{-i(\omega_-+\alpha\kappa_-)t}E_+^*
        \nonumber \\&&         +e^{+i(\omega_-+\alpha\kappa_-)t}E_+).
\end{eqnarray}
\end{mathletters}
Note that fluctuations in the lasing mode $E_-$ or the total density $N$
do not appear in the linearized equations for $E_+$ and $\Delta$. 
The fluctuations in the non-lasing mode are therefore not correlated with
the fluctuations in the lasing mode.

The laser field itself does affect the dynamics of fluctuations, however,
as expressed by the final terms of the linearized equations which 
are sensitive to both the amplitude $\sqrt{I_0}$ and the phase
$-(\omega_-+\alpha\kappa_-)t$ of the laser field. In order to describe the 
phase relation between the fluctuations of the non-lasing mode $E_+$ and the 
laser field, $E_+$ can be expressed
in terms of the component $f_\parallel$ in phase with the laser field and 
the component $f_\perp$ which is $\pi/2$ out of phase with the laser field,
\begin{eqnarray}
E_+ &=& (f_\parallel-i f_\perp)e^{-i(\omega_-+\alpha\kappa_-)t}.
\end{eqnarray}
Using this parameterization the Langevin equation describing the spontaneous 
emission from the non-lasing mode formulated as a matrix equation reads
\footnotesize
\[
\frac{d}{dt}
\left(\begin{array}{c}f_\parallel\\f_\perp\\ \Delta\end{array}\right)
=
\]
\begin{equation}
\left(\begin{array}{ccc}
        -s           & +\Omega &       \frac{w\sqrt{I_0}}{2}\\
     -\Omega         &    -s   & \alpha\frac{w\sqrt{I_0}}{2}\\
-4\kappa_-\sqrt{I_0} &     0   &  -\gamma-2\Gamma-wI_0      
\end{array}\right)               
\left(\begin{array}{c}f_\parallel\\f_\perp\\ \Delta\end{array}\right)
+
\left(\begin{array}{c}Q_\parallel\\Q_\perp\\ 0\end{array}\right),
\end{equation}
\normalsize
where $\Omega=\omega_--\omega_+$ and $s=\kappa_+-\kappa_-$ are the frequency 
and loss differences between the modes and $Q_\parallel$ and $Q_\perp$ are the
quantum noise terms acting on the cavity mode from the quantum vacuum outside
the cavity and from the dipole fluctuations in the gain medium. Pump noise
in the carrier density has been neglected since its effect is much smaller
than that of the quantum fluctuations in the case of strong relaxation 
oscillations investigated in this letter. The minimal quantum fluctuations 
for complete inversion in the gain medium are given by
\begin{equation}
\langle Q_\parallel(t)Q_\parallel(t+\tau) \rangle =
\langle Q_\perp(t)Q_\perp(t+\tau) \rangle = 
\kappa\delta(\tau).
\end{equation}
The fluctuation dynamics depends on four timescales, the damping coefficients
of the field $s$ and of the carrier density difference $\gamma+2\Gamma+wI_0$, 
the frequency difference between the supermodes $\Omega$ and the coupling 
frequency between the field and the carrier density difference
$\sqrt{2\kappa_-wI_0}$. Note that this coupling frequency is identical
with the frequency of relaxation oscillations in the total intensity.
Physically, it represents the spatial variations in the relaxation 
oscillations associated with fluctuating spatial hole-burning.
If the spatial hole-burning effect represented by $\Delta$ is negligible,
the fluctuations in $f_\parallel$ and $f_\perp$ are rotations with frequency
$\Omega$ damped at a rate of $s$. This corresponds to a single Lorentzian
spontaneous emission line at $\omega_+$ with a width of $2s$. 

We will now examine the opposite case of strong spatial hole-burning effects, however, which 
applies if the frequency difference $\Omega$ is much smaller than the 
relaxation oscillation frequency $\sqrt{2\kappa_-wI_0}$. In this limit, the light field is
phase locked to the laser field by the spatial hole-burning associated with
the in-phase field fluctuations $f_\parallel$. The field dynamics separates 
into two uncorrelated solutions. The out-of-phase fluctuations are described 
by an exponential relaxation of fluctuations in $f_\perp$,
\begin{eqnarray}
\Delta &\approx& f_\parallel \approx 0 \nonumber \\
f_\perp &\approx& f_0 e^{-(s+\alpha\Omega)t}.
\end{eqnarray}
The in-phase fluctuations are actually phase shifted with respect to the 
laser light by the linewidth enhancement factor $\alpha$. They are given by
relaxation oscillations with
\begin{eqnarray}
\label{eq:fluct}
f_\parallel&\approx&f_0 \cos (\sqrt{2\kappa_-wI_0}t) 
                e^{-(\gamma+2\Gamma+wI_0+s-\alpha\Omega)t/2} \nonumber \\
f_\perp&\approx&\alpha f_0 \cos (\sqrt{2\kappa_-wI_0}t) 
                e^{-(\gamma+2\Gamma+wI_0+s-\alpha\Omega)t/2} \nonumber \\
\Delta &\approx& -2\sqrt{\frac{2\kappa_-}{w}}f_0 \sin (\sqrt{2\kappa_-wI_0}t)
\nonumber \\ && e^{-(\gamma+2\Gamma+wI_0+s-\alpha\Omega)t/2}.
\end{eqnarray}   
The complete solution of the Langevin equation and the resulting two-time
correlation functions are analogous to the ones derived for
the polarization fluctuations in VCSELs \cite{Hof97}. 

The spectrum of the spontaneous emission into the non-lasing mode is a
triplet composed of a central line at the laser frequency 
representing the out-of-phase emission processes and two sidebands representing 
the relaxation oscillations caused by in-phase spontaneous emission. 
The total spectrum reads
\[
I(\delta\omega)=
   \frac{\frac{\kappa_-}{2\pi}(1+\alpha^2)}
        {(s+\alpha\Omega)^2+\delta\omega^2}
\]
\[
+\frac{\frac{\kappa_-}{2\pi}(1+\alpha^2)}     
        {(\gamma+2\Gamma+wI_0+s-\alpha\Omega)^2+
         4(\delta\omega-\sqrt{2\kappa_-wI_0})^2}
\]
\begin{equation}
+\frac{\frac{\kappa_-}{2\pi}(1+\alpha^2)}
        {(\gamma+2\Gamma+wI_0+s-\alpha\Omega)^2+
         4(\delta\omega+\sqrt{2\kappa_-wI_0})^2}.
\end{equation}
The Intensity is given in terms of the average photon number inside the 
optical cavity. The emission rate is found by multiplying with $2\kappa_+$.
$\delta\omega$ is the frequency difference to the position of the laser
line at $\omega_-+\alpha\kappa_-$.
Figure \ref{triplet} shows the spontaneous emission triplet of the non-lasing supermode
for a typical choice of parameters, $\gamma = 10$ GHz, $\Gamma = 20$ GHz,
$\kappa_-=2000$ GHz, $\alpha = 3$, $\Omega = 5$ GHz and $s=15$ GHz. The 
laser intensity is 
given in units of $\gamma/w$, which typically corresponds to about 
10000 photons in the cavity. 

The central line does not change as the laser intensity increases. The reason 
lies in the fact that the interference term of the out-of-phase fluctuations 
and the laser light vanishes. The spatial intensity distribution is not 
modified by out-of-phase spontaneous emission and no spatial hole-burning 
results. Therefore, this component of the spontaneous emission is the unchanged emission from the
carrier density $N$ pinned at $N=2\kappa_-/w$. 

The sidebands at $\delta\omega = \pm \sqrt{2\kappa_-wI_0}$ represent that part of the
spontaneous emission which has a component in phase with the laser light.
The non-zero interference term between the non-lasing mode and the
laser light causes spatial hole-burning as represented by the parameter 
$\Delta$. Consequently, there are spatial relaxation oscillations at the
relaxation oscillation frequency $\sqrt{2\kappa_-wI_0}$. Since the Interference
term between the non-lasing mode and the laser light represents the intensity
difference between the two lasers in the array, the oscillation in the
in-phase component of the non-lasing field $f_\parallel$ corresponds to
a spatial intensity oscillation.
The phase of the sideband emission processes is not the same as the phase of the laser
field, however. As can be seen in equation (\ref{eq:fluct}), the out-of-phase
component $f_\perp$ is $\alpha$ times as large as the in-phase component
$f_\parallel$. This is a consequence of the spatial change in the refractive
index. While the gain-guiding properties associated with a non-zero carrier
density difference $\Delta$ convert the field of the lasing supermode into
the non-lasing supermode at the same phase, the index guiding properties
change the phase by $\pi/2$.  
Since the ratio of the index guiding and the gain guiding induced by the
changes in carrier density is given by $\alpha$, the phase difference between
the laser field and the relaxation oscillation sidebands of the non-lasing
mode is equal to $\arctan (\alpha)$. 

Since the relaxation oscillation sidebands represent the spatial effects 
induced
by the in-phase component of spontaneous emission into the non-lasing mode, 
the linewidth, the frequency, and the total intensity are a 
function of the carrier dynamics as well as of the laser field.
The dependence on laser intensity is shown in figure \ref{sides}, in which 
the out-of-phase
contribution has been removed. The total intensity in the two sidebands 
$I_{sb}$ compared to the total intensity of the central line $I_{cl}$ is
\begin{equation}
\frac{I_{sb}}{I_{cl}} = \frac{s+\alpha\Omega}
{\gamma+2\Gamma+wI_0+s-\alpha\Omega}.
\end{equation}
The diffusion $\Gamma$ and the total rate of induced and spontaneous 
transitions $\gamma+wI_0$ both suppress spontaneous emission processes into the 
non-lasing supermode by damping the carrier density difference $\Delta$.
In this manner, the total spontaneous emission into the non-lasing supermode
decreases as laser intensity increases, even though the average carrier
density remains pinned at $N=2\kappa_-/w$. 

The spontaneous emission triplet of the non-lasing supermode should be
observable in the spectrum emitted from the symmetric non-lasing 
supermode of the laser array cavity. Since the symmetric
supermode has its far-field intensity maximum in the center of the far-field,
where the stable anti-symmetric laser mode should have zero intensity, it
seems likely, that this spectrum can be observed by measurements of the weak
intensities in the center of the far field.
The assumptions of the model applied in this letter may then be tested
by measurements of the linewidths and frequencies. An interesting point 
would be the comparison of sideband intensities. If the coherent coupling
$\Omega$ is such that $\sqrt{2\kappa_-wI_0} \gg \Omega$ does not hold anymore, then the lower
frequency sideband intensity should be stronger than the high frequency sideband intensity -- up to the 
point where $\Omega \gg \sqrt{2\kappa_-wI_0}$ and there is a single spontaneous
emission line left at $\omega_+$. This case would imply extremely strong 
coupling between the individual lasers of the array, however, implying
that the array acts as a single laser with a modified cavity. 

In conclusion, we have shown that the carrier dynamics of a two laser array 
modifies the spontaneous emission in the non-lasing supermode by phase locking
it to the laser field and by modulating the in-phase component through 
relaxation oscillations of the spatial hole-burning. 

%
%
%
\newpage
\vspace*{1cm}
\begin{figure}
\caption{Spontaneous emission triplet for $\gamma = 10$ GHz, $\Gamma = 20$ GHz,
$\kappa_-=2000$ GHz, $\alpha = 3$, $\Omega = 5$ GHz and $s=15$ GHz. (a) shows
the dependence of the frequencies, the intensities and the linewidths on the
laser intensity as a three dimensional plot and (b) shows the triplet at 
Intensities of $wI_0 = 2.5$ GHz, $I_0=5$ GHz, $I_0=7.5$ GHz and $I_0=10$ GHz.}
\label{triplet} 
\end{figure}
\begin{figure}
\caption{Sidebands without the out-of-phase line at $\delta\omega = 0$.
As the intensity increases, the in-phase contribution of spontaneous 
emission is suppressed by the stimulated emission term $wI_0$.}
\label{sides} 
\end{figure}
%
%
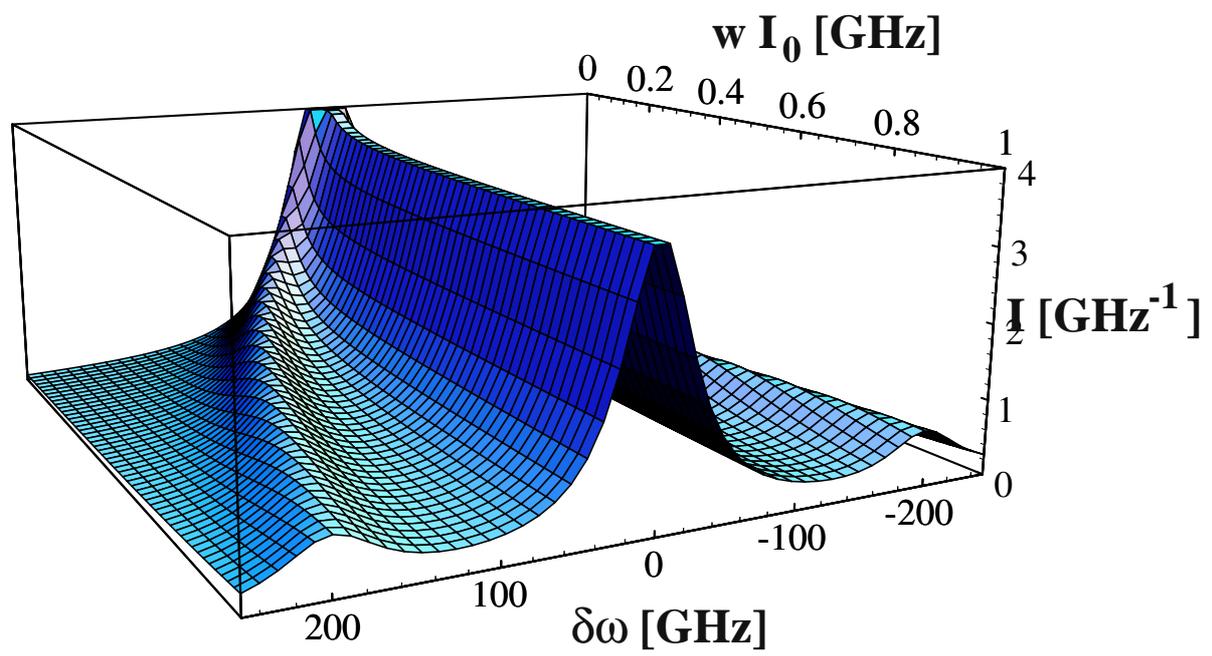
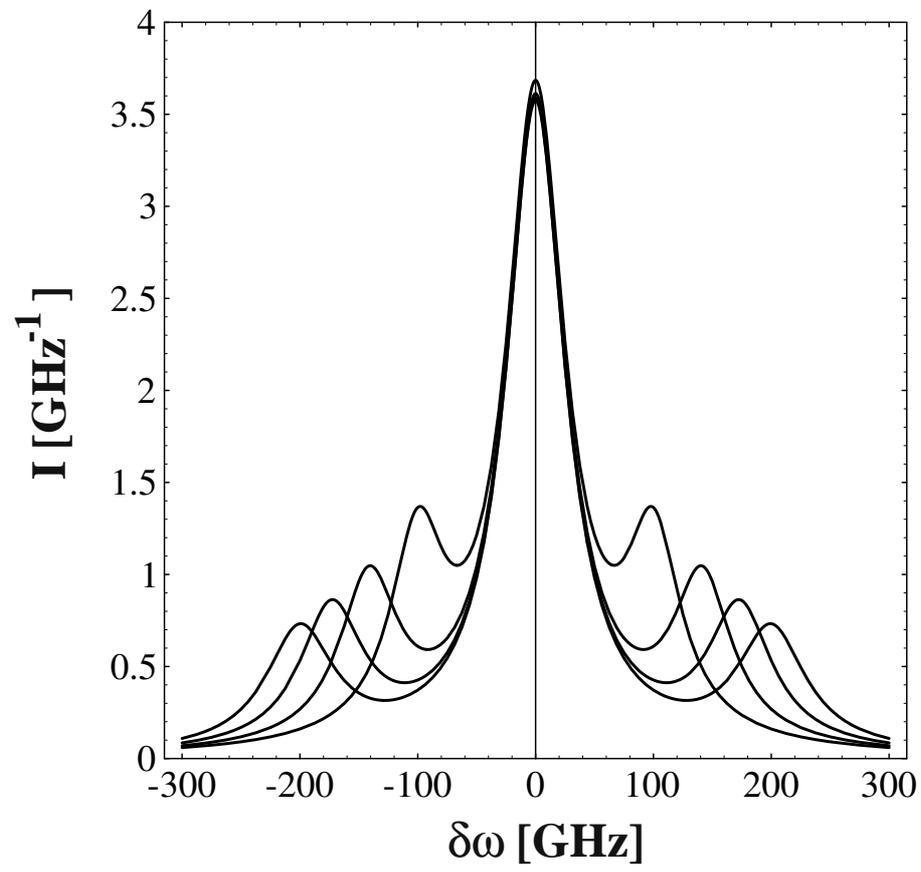
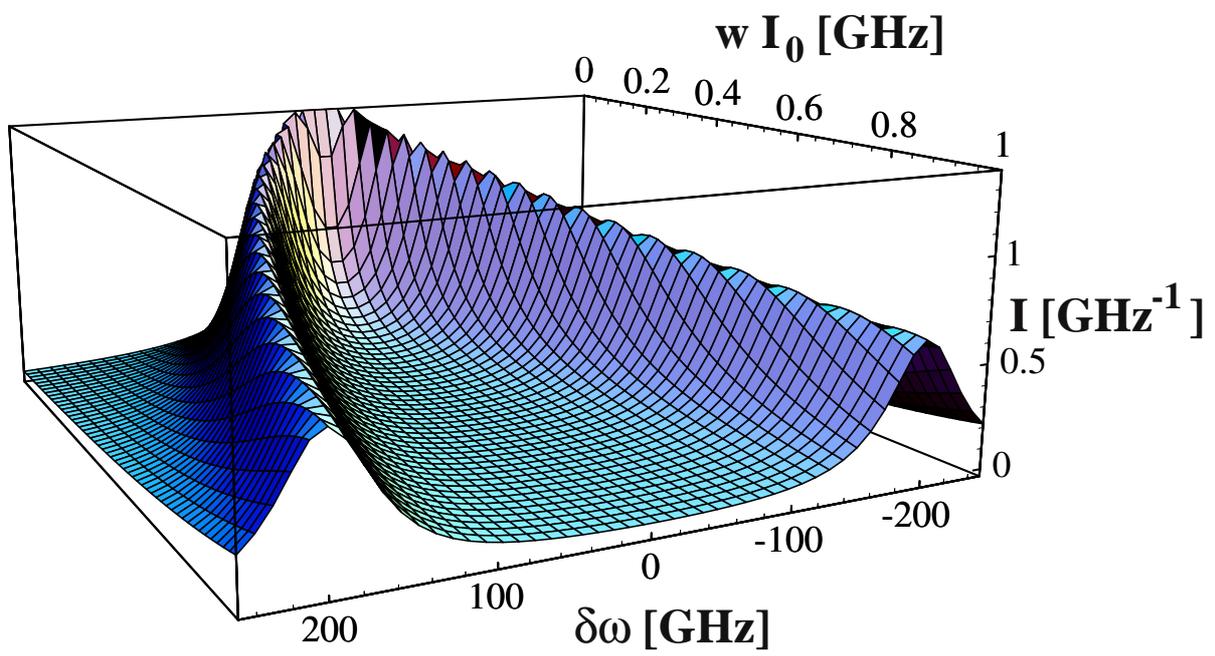
\end{document}